\documentclass[11pt]{article}

\usepackage[T1]{fontenc}
\usepackage{lmodern}
\usepackage{microtype}
\usepackage{geometry}
\everymath{\displaystyle}
\usepackage{mathtools}
\usepackage{amsmath,amsfonts,amssymb,amsthm}
\usepackage{graphicx}
\usepackage{epstopdf} 

\usepackage[authoryear,round]{natbib}
\usepackage{authblk}
\usepackage[hidelinks]{hyperref}

\begin{document}
\title{Market Making and Transient Impact in Spot FX}

\author{Alexander Barzykin\thanks{Email: \texttt{alexander.barzykin@hsbc.com}}}
\affil{HSBC, 8 Canada Square, Canary Wharf, London E14 5HQ, United Kingdom}
\date{January 17, 2026}

\maketitle
\begin{abstract}
\noindent
Dealers in foreign exchange markets provide bid and ask prices to their clients at which they are happy to buy and sell, respectively. In order to manage risk, dealers can skew their quotes and also hedge in the interbank market. Hedging offers certainty but comes with transaction cost and market impact. Optimal market making with execution has previously been addressed within Almgren-Chriss market impact model, with instantaneous and permanent components. However, there is an overwhelming empirical evidence of the transient nature of market impact, with instantaneous and permanent impact arising as two limiting cases. In this note, we consider an intermediate scenario and study the interplay between risk management and impact resilience.
\end{abstract}

\vspace{7mm}
\vspace{5mm}

\section{Introduction}

Foreign exchange (FX) markets continue to operate largely on OTC (over-the-counter) basis where dealers offer bid and ask prices to their clients bilaterally or via aggregators.
The dealer aims to make spread but has to manage inventory risk arising due to asynchronicity of client flow and market volatility.
In order to attract risk-reducing flow, the dealer can skew their quotes but ultimately may chose to hedge excess inventory in the interbank market.
Hedging offers certainty but comes with transacation cost and market impact.
Optimal strategies balancing spread capture and risk management have been a subject of recent active research \citep{avellaneda2008,gueant2013,cartea2014,butz2019,bergault2021b,barzykin2023,barzykin2025a}.
In particular, internalization versus extenalization dilemma has been in focus \citep{butz2019}, and Almgren-Chriss \citeyearpar{almgren2001} model with instantaneous cost and linear permanent impact has been employed to describe the execution in the interbank market \citep{barzykin2023}.
One of the conclusions of this research is the existence of the pure internalization zone where it is not optimal for the dealer to execute.
The inventory threshold, beyond which the dealer will start to execute, depends on risk aversion, volatility, client flow and, importantly, transaction cost and market impact.
There is overwhelming empirical evidence of the transient nature of market impact with the propagator model of \citet{bouchaud2018} capturing the essence of the phenomenon.
A particular case of exponential relaxation in Obizhaeva-Wang model \citeyearpar{obizhaeva2013} comes with a clear explanation in terms of the limit order book resilience.
At the same time, large order execution demonstrates a universal square root dependence on the total executed quantity \citep{toth2011,sato2025}.
Both effects are found to be very important in optimal execution \citep{neuman2022,hey2023,webster2023}.
So, why Almgren-Chriss?

First of all, the square root law is a meta feature which arises for relatively large parent orders due to latent liquidity \citep{bucci2019} and/or sophisticated traders able to take advantage of the market over-reaction to the metaorder \citep{durin2023}.
For smaller sizes/participation rates, the impact is found to be linear \citep{bucci2019,durin2023}.
The primary route of dealer's risk reduction is internalization, so linearity of hedging impact is a natural assumption, ensuring no dynamic arbitrage \citep{gatheral2010}.
Secondly, the impact, albeit transient, is know to be very persistent \citep{bouchaud2018} while the typical timescale of risk relaxation for FX dealers is short \citep{butz2019}.
Thus, the assumption of permanent impact (on the timescale of risk relaxation) is also plausible.
However, one can reasonably conjecture that at least part of the impact decay should be comparable to the risk decay, 
as execution trades can end up in other market maker's inventory, and there is indeed evidence of fast impact decay in spot FX \citep{schmidt2016}.
In this scenario, the interplay between the impact resilience and risk management can become important.
\citet{cartea2026} have recently studied the Nash equilibrium between a dealer and two clients -- an informed and an unformed trader, in continuous setting, incorporating the dealer's 
hedging activity in the lit market with exponential market impact kernel.
They have observed the fundamental change in the trajectories of the inventory of the informed trader depending on the value of the decay parameter and found a range of values for which it may be beneficial for the trader to ride the price impact.

We begin by demonstrating that OTC trading indeed implies the propagator type of impact, as in Obizhaeva-Wang \citeyearpar{obizhaeva2013} and echoing the findings of \citet{eisler2016} in a different market.
Then we introduce the resilient impact state into the market making model and solve the corresponding Hamilton-Jacobi-Bellman (HJB) equation to find optimal controls both numerically and analytically, within a quadratic approximation à la \citet{bergault2021a}.
Using parameters relevant to institutional FX, we find visible benefit in performance when risk managing large inventories and taking resilient impact into account.

\section{Transient Impact of Client Trades with the Dealer}

Let us demonstrate that the market making model itself implies a transient impact, namely, as soon as the client trades with the market maker, the price jumps and then relaxes back to its initial value.
We consider a standard Avellaneda-Stoikov \citeyearpar{avellaneda2008} OTC market making model with Bergault-Guéant \citeyearpar{bergault2021b} extension to a ladder of sizes.
The dealer offers bid/ask prices $S_t^{n,b/a} = S_t \mp \delta_t^{n,b/a}$ for a set of trade sizes $\Delta^n$ ($0<\Delta^1<\ldots<\Delta^N$, $N\ge1$) to clients.
Here the mid-price $S_t$ is modeled as a Brownian motion and deltas are dealer's controls to optimize risk-adjusted P\&L over a finite time horizon $T$ \citep{cartea2014}.
We consider Poisson trade arrivals with side-symmetric intensity $\lambda^{n, b/a}(\delta) = \lambda^n (\delta)$.
This formulation is standard and the corresponding baseline HJB equation for the value function $V(t, q)$ at time $t$ and inventory $q$ reads \citep{gueant2013,cartea2015,gueant2016}
\begin{equation}
0 = \partial_t V - \frac{\gamma \sigma^2 q^2}{2} + \sum_n \Delta^n \Bigg(
H^n_\text{\tiny{OTC}}\left[ D_{q+}^n V(t,q) \right]
+ H^n_\text{\tiny{OTC}}\left[ D_{q-}^n V(t,q) \right]
\Bigg)
\end{equation}
with $t \in [0, T]$ and appropriate terminal condition
\footnote{Terminal penalty does not alter optimal controls at time 0 when $T$ is sufficiently large, 
so one can assume $V(T,q) = 0$.}
at $t=T$, where $\sigma$ denotes volatility, $\gamma$ the risk aversion coefficient and
\begin{equation}
D_{q\pm}^n V(t,q) = \frac{V(t, q) - V(t, q \pm \Delta^n)}{\Delta^n}
\end{equation}
\begin{equation}
H^n_\text{\tiny{OTC}}(p) = \sup_{\delta \in \mathbb{R}} \lambda^n (\delta) (\delta-p)
\end{equation}

Following \citet{bergault2021a} we seek an approximate solution via second order Hamiltonian expansion and quadratic ansatz for the value function,
leading to Riccati ODE.
Thus, we write
\footnote{Linear term disappears due to symmetry.}
$V(t,q) = - A(t) q^2 - C(t)$ and obtain a stationary solution:
\footnote{$C(t)$ does not influence the controls and thus not shown.}
\begin{equation}
\label{a0}
A_0 = \lim_{T \to \infty} A(0) = \sigma \sqrt{\frac{\gamma}{8\xi}}, \qquad \xi = \sum_n H^{n''}_\text{\tiny{OTC}}(0) \Delta^n
\end{equation}
where 
\begin{equation}
H^{n''}_\text{\tiny{OTC}}(0) = \frac{\lambda^n(\delta^n_0)}{\delta^n_0 c^n}, \qquad
c^n = 2 - \frac{\lambda^n(\delta^n_0) \lambda^{n''}(\delta^n_0)}{(\lambda^{n'}(\delta^n_0))^2} 
\end{equation}
\begin{equation}
\delta^n_0 = \underset{\delta \in \mathbb{R}}{\text{argmax}}\, \delta \lambda^n (\delta).
\end{equation}
Optimal quotes are, therefore,
\begin{equation}
\delta_*^{n,b/a} \simeq  \delta^n_0 + \frac{A_0}{c^n} \left(\Delta^n \pm 2q\right)
\end{equation}
where $\pm$ corresponds to $b/a$, respectively.
This approximate solution is well-known, even though it may not have appeared explicitly in the early literature.
It provides a valid expansion in the limit of low risk aversion and can serve as a starting point for policy iteration -- an efficient method for solving HJB numerically.\\

We can immediately see an instantaneous impact signature in optimal quotes.
The equilibrium cost (half-spread) for size $\Delta^n$ is $\delta^n_0 + A_0\Delta^n/c^n$ but as soon as one trades, the mid-price jumps by $2A_0 \Delta^n/c^n$ ensuring no-arbitrage.\footnote{The weighted average price of executing two consecutive $\Delta^n$ orders should equal the price of a single $2\Delta^n$ order.}
Under stationary control, the ODE for the mean-field inventory $\bar{q}_t = \mathbb{E}[q_t]$  reads
\begin{equation}
\dot{\bar{q}}_t = \sum_n \Delta^n \Big(\lambda^{n}\big(\delta^{n,b}_*(\bar{q}_t)\big) - \lambda^{n}\big(\delta^{n,a}_*(\bar{q}_t)\big)\Big) \simeq - \omega \bar{q}_t
\end{equation}
where
\begin{equation}
\omega = \sigma \sqrt{2\gamma \xi}
\end{equation}
which is exponential relaxation. After an inventory shock of magnitude $q_0$, the dealer's position relaxes as $\bar{q}_t = q_0 \exp (-\omega t)$, and so does optimal pricing.
This is essentially an OTC equivalent of the Obizhaeva-Wang \citeyearpar{obizhaeva2013} model.
While this is understandably an internal transient impact on a particular dealer's pricing, skewing is known to potentially influence the external market as well via price reading \citep{barzykin2025a}, not to mention direct market impact of the external hedging activity.
Note that similar arguments are applicable to passive OTC impact \citep{barzykin2025b}, where the dealer skews in anticipation of the flow to fill a passive client order rather than in response to an inventory change due to a client trade.
Although the impact amplitude is different in this case, influenced by a number of factors including the dealer's expectation of the properties of the passive flow as well as on any flow facilitation arrangements, the relaxation time still originates from the franchise flow assymetry driven by the skew.

\section{Hedging with Transient Impact}

Let us now return to market maker's optimization problem.
In addition to skewing, the dealer can trade in the interbank market to hedge their risk.
As in \citet{barzykin2023}, it is assumed that hedging is continuous with speed $v_t$ and transaction cost $L(v) = \psi |v| + \eta v^2$.
Hedging also causes market impact which decays exponentially with time, as in Obizhaeva-Wang \citeyearpar{obizhaeva2013}.
Thus, we write for the mid-price evolution
\begin{equation}
dS_t = \sigma dW_t + dx_t, \qquad dx_t = (-\beta x_t + k v_t) dt
\end{equation}
where $x_t$ denotes the resilient impact state.
Clearly, $\beta = 0$ brings us back to Almgren-Chriss \citeyearpar{almgren2001}.
The corresponding HJB for the value function $V(t,q,x)$ reads (after standard substitution to eliminate price diffusion):
\begin{equation}
0 = \partial_t V - \beta x (q + \partial_x V) - \frac{1}{2}\gamma \sigma^2 q^2 
+ H_\text{\tiny{E}}(p_\text{\tiny{E}})
+ \sum_n \Delta^n \Bigg(
H^n_\text{\tiny{OTC}}\left[ D_{q+}^n V \right]
+ H^n_\text{\tiny{OTC}}\left[ D_{q-}^n V \right]
\Bigg)
\end{equation}
with zero terminal condition and
\begin{eqnarray}
&&H_\text{\tiny{E}}(p) = \sup_{v \in \mathbb{R}} pv - L(v) = \frac{(|p|-\psi)^2_+}{4\eta}, \\
&&p_\text{\tiny{E}} = k(q + \partial_x V) + \partial_q V
\end{eqnarray}
The terms $-\beta x (q + \partial_x V)$ come from the controlled drift in $S$ and the state drift of $x$.
The optimal quotes are given by (using envelope theorem)
\begin{equation}
\label{oquotes}
\delta_*^{n,b/a} = (\lambda^n)^{-1} \left( -H^{n'}_\text{\tiny{OTC}}\left[ D_{q\pm}^n V(t,q,x)\right]\right)
\end{equation}
and the optimal execution speed is
\begin{equation}
\label{ospeed}
v_* = \begin{cases}
\mathrm{sign}(p_\text{\tiny{E}}) \frac{|p_\text{\tiny{E}}| - \psi}{2\eta}, 
& \text{if $|p_\text{\tiny{E}}| > \psi$}.\\
0, & \text{otherwise}.
\end{cases}
\end{equation}

For qualitative insight let us again follow \citet{bergault2021a} and approximate $H^n_\text{\tiny{OTC}}(p)$ up to the second order in $p$
and then use quadratic ansatz for the value function
\footnote{Linear terms vanish due to symmetry and the $x^2$ term is small.}
\begin{equation}
V(t,q,x) = -A(t) q^2 - B(t) q x - C(t)
\end{equation}
The execution Hamiltonian $H_\text{\tiny{E}}(p)$ is neglected when approximating the value function due to spread cost.
\footnote{One can neglect spread and keep quadratic cost as an alternative approximation but the difference is insignificant.}
HJB thus reduces to the Riccati system for the coefficients and
the stationary solution is given by
\footnote{The free term does not influence the controls and thus not shown.}
\begin{equation}
B_0 = \lim_{T \to \infty} B(0) = \frac{\beta}{\beta + \omega}
\end{equation}\\
with $A_0$ being the same as defined earlier in Eq.~(\ref{a0}).
$B_0$ directly reflects the competition between impact resilience and risk relaxation via client flow.
Optimal quotes can be obtained from Eq.~\ref{oquotes}.
Keeping only the first significant correction we obtain
\begin{equation}
\delta_*^{n,b/a}(q, x) \simeq \delta^n_0 + \frac{A_0}{c^n}(\Delta^n \pm 2q ) \pm \frac{B_0}{c^n} x
\end{equation}
with additional linear dependence on the impact state.
Optimal execution speed is driven by the value of $p_\text{\tiny{E}}$ which is
now given by
\begin{equation}
\label{pexe}
p_\text{\tiny{E}}(q,x) = -\big(2A_0 - k(1-B_0)\big) q - B_0 x
\end{equation}
$|p_\text{\tiny{E}}(q,x)| = \psi$ defines the boundary of the pure internalization zone.

\begin{figure}[h]
\centering
\includegraphics[width=0.95\columnwidth]{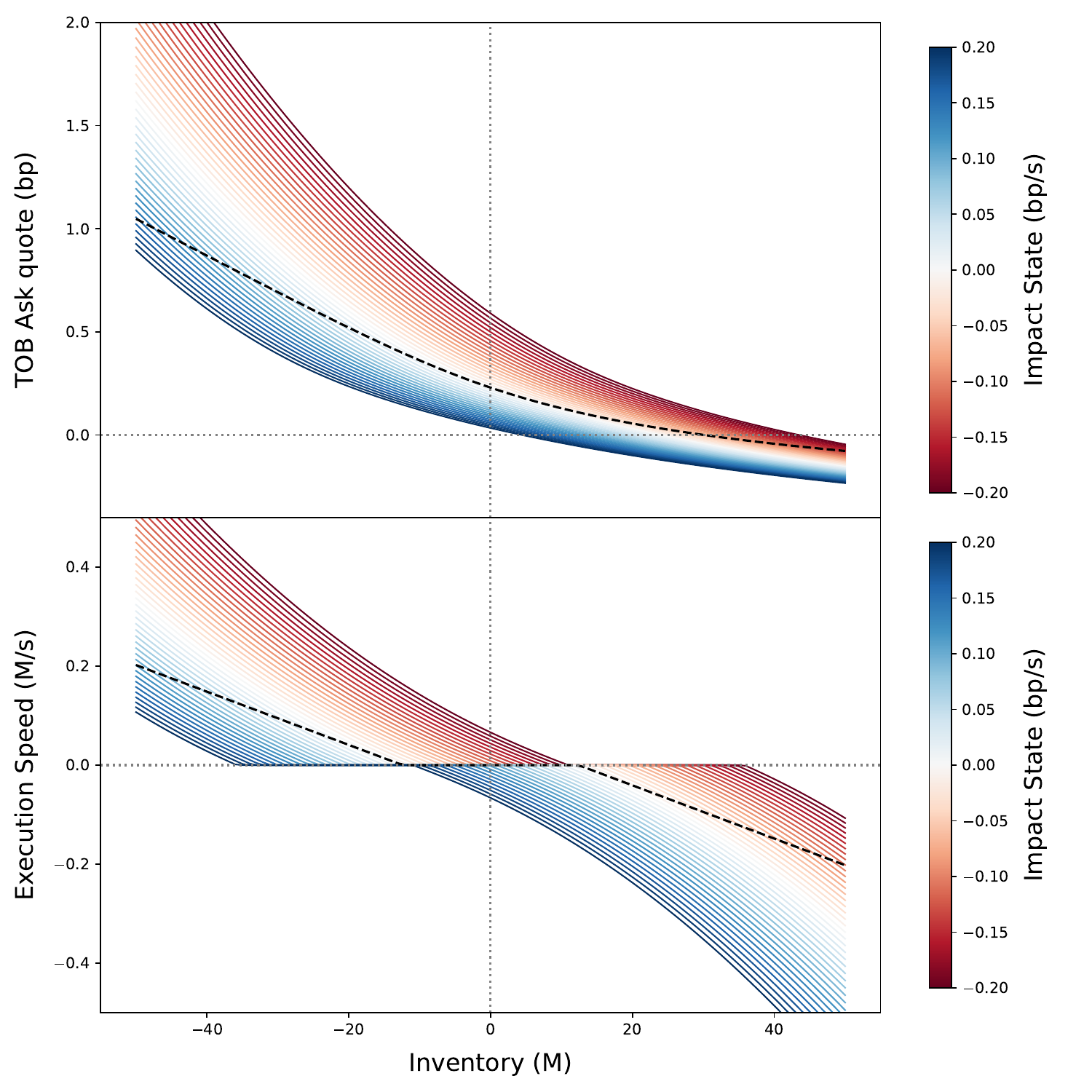}
\caption{
Optimal top of book (TOB) ask quote $\delta^{1,a}_*$ and execution speed $v_*$ as functions of inventory $q$ and impact state $x$ for a set of parameters defined in the text.
Dashed line corresponds to approximate solution for $x=0$.
}
\label{optimal}
\end{figure}

\section{Numerical Examples}

As an illustration, consider a standard size ladder of $\Delta^n = (1, 2, 5, 10, 20, 50)$ M notional and a sigmoid intensity function $\lambda^n(\delta) = \lambda^n_0 \big(1 + \exp(a^n + b^n \delta)\big)^{-1}$ 
with amplitudes $\lambda_0^{n} = (2000, 800, 600, 400, 100, 50)$~day$^{-1}$ and uniform parameters $a^n = -1$, $b^n = 7$ bp$^{-1}$.
Here bp stands for basis points.\footnote{This implies GBM while we deal here with simple Brownian motion. 
The difference is negligible in FX market making due to short trading horizons.}
This set of parameters corresponds to a liquid currency pair with a daily turnover of $ \simeq 5$ billion notional and a top-of-book spread of $\simeq 0.5$~bp. 
We also assume a daily volatility of $100$~bp and a risk aversion coefficient of $\gamma = 10^{-3}$~bp$^{-1}$~M$^{-1}$.
Execution related parameters are $\psi = 0.2$~bp, $\eta = 1.5$~bp$\cdot$s/M, $k = 0.005$~bp/M and $\beta = 1000$~day$^{-1}$.
The impact decay rate was intentionally chosen to be comparable to the risk relaxation time ($\omega \simeq 560$~day$^{-1}$ in this case).

\begin{figure}[h]
\centering
\includegraphics[width=0.95\columnwidth]{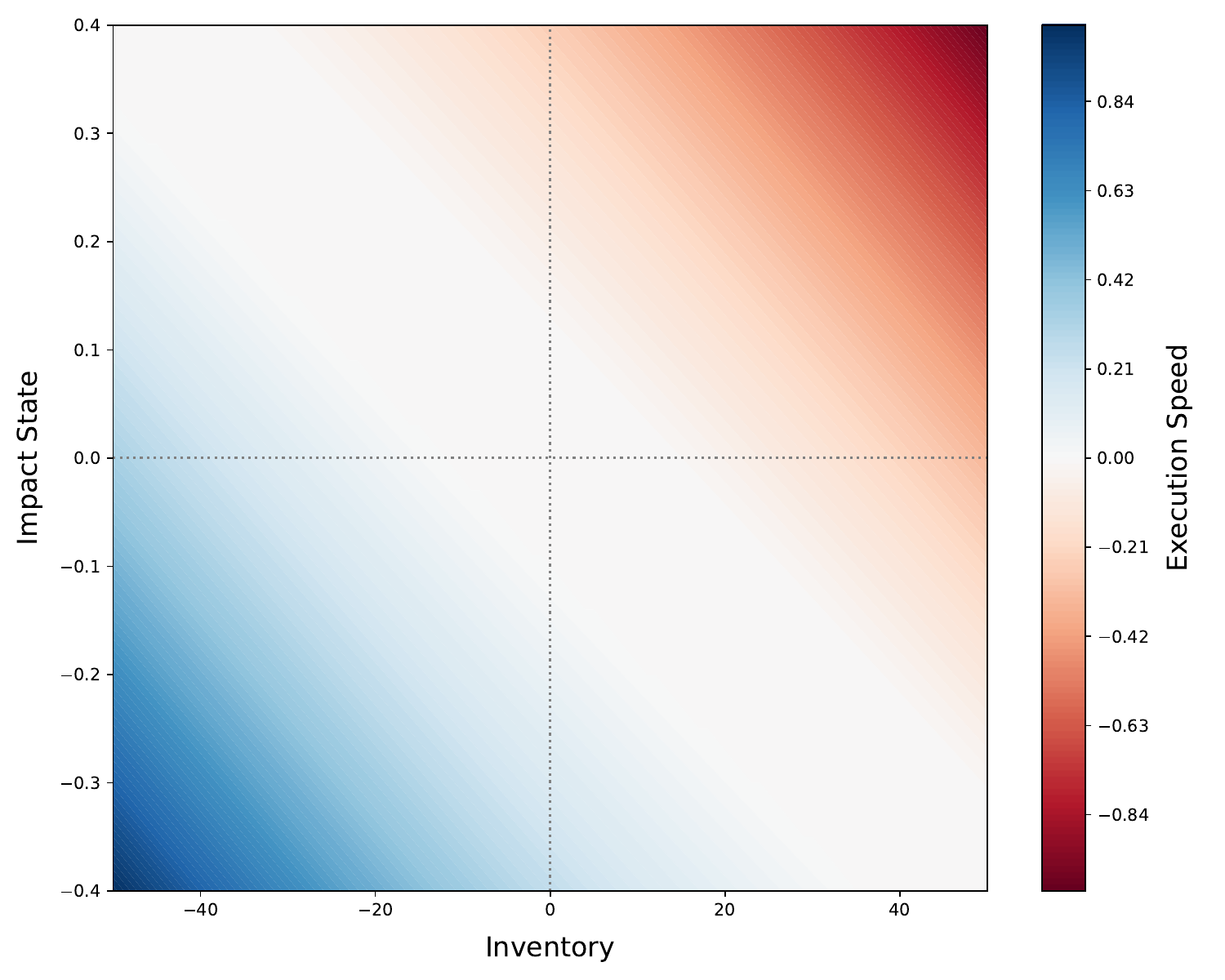}
\caption{
2d plot of optimal execution speed $v_*$ as a function of inventory $q$ and impact state $x$ for a set of parameters defined in the text.
}
\label{execution2d}
\end{figure}

Figure~\ref{optimal} illustrates the optimal controls $\delta^{1,a}_*$ and $v_*$ as functions of the dealer's inventory $q$ and the impact state $x$ (colour coded).
The solution was obtained by numerically solving HJB using explicit Euler scheme (taking sub-second in jax).
The overall shape looks similar to the optimal solution with Almgren-Chriss impact \citep{barzykin2023} but the dependence on the impact state is significant.
In particular, pure internalization zone depends on the impact state, as also shown in Figure~\ref{execution2d}, which is understandable because impact relaxation is a kind of price prediction.
Approximate closed-form solution, obtained by substituting quadratic expansion of the value function into Eqs.~(\ref{oquotes}) and (\ref{ospeed}), is found to capture the optimal solution qualitatively but deviates for larger inventories, as expected.

\begin{figure}[h]
\centering
\includegraphics[width=0.95\columnwidth]{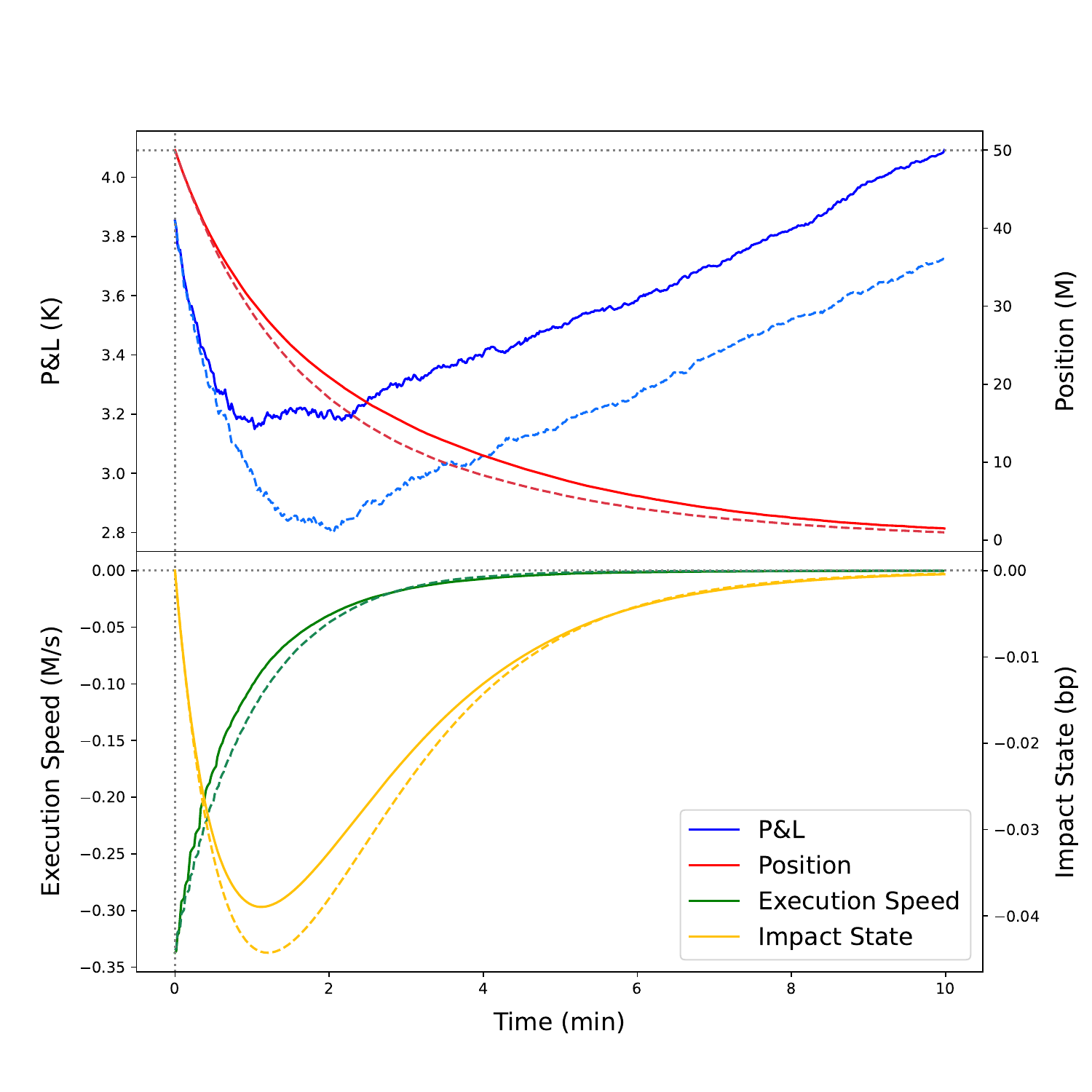}
\caption{
P\&L, position, execution speed and impact state dynamics following an inventory shock of $q_0 = 50$~M for a parameter set defined in the text.
The results were obtained via Monte Carlo simulation with $10^4$ trajectories.
Dashed lines correspond to the model optimized with Almgren-Chriss impact.
}
\label{relaxation}
\end{figure}

Figure~\ref{relaxation} shows how P\&L, position, execution speed and impact state evolve following an inventory shock (large client trade).
The results were obtained via Monte Carlo simulations with $10^4$ trajectories using the corresponding SDE and optimal controls (numerically exact).
As expected, we can see the inventory decreasing due to skewing and hedging with the absolute execution speed also decreasing together with the inventory, while the absolute impact goes through a maximum.
P\&L jumps at time zero as the dealer makes half spread, then it decreases due to impact but then rises again due to continuing flow monetization.
We compare the performance of optimal controls with transient impact against those with permanent impact (i.e., HJB is solved with $\beta=0$) while in both cases the actual underlying impact is transient.
The benefit of incorporating impact resilience is very clear in this case.
However, if large trades are rare, the majority of client flow will be internalized, so that the impact state will rarely be sufficiently large to cause a significant difference to the total expected P\&L.
Therefore, Almgren-Chriss model can be a reasonable approximation after all.
Nevertheless, risk management of larger trades is usually under scrutiny by the desk, and this is where taking into account impact resilience becomes important.

\section*{Acknowledgment}
The author would like to express his sincere gratitude to Olivier Guéant (Université Paris Cité) and Leandro S{\'a}nchez-Betancourt (University of Oxford) for fruitful discussions and valuable comments and to Richard Anthony (HSBC) for support throughout the project.
The views expressed are those of the author and do not necessarily reflect the views or the practices at HSBC.

\end{document}